\documentclass{aa}  
\usepackage{graphicx}
\usepackage{txfonts}
\usepackage{xcolor}
\usepackage{url}
\urldef\myurl\url{https://archive.eso.org/scienceportal/home?data_release_date=*:2019-09-18&pos=39.2442,-34.57798&r=0.016667&poly=39.318719,-34.614387,39.169781,-34.614387,39.169846,-34.541445,39.318654,-34.541445&dp_type=SPECTRUM&sort=dist,-fov,-obs_date&s=P%2fDSS2%2fcolor&f=0.122541&fc=39.318654,-34.541445&cs=J2000&av=true&ac=false&c=8,9,10,11,12,13,14,15,16,17,18&ta=RES&dts=true&sdtm=%7b%22SPECTRUM%22%3atrue%7d&at=39.2442,-34.57798}
\usepackage[pdftex]{hyperref}
\hypersetup{colorlinks=true,linkcolor=black,citecolor=blue,urlcolor=blue,draft}

\newcommand{\kepler}{\textit{Kepler}\xspace}
\newcommand{\tess}{TESS\xspace}
\newcommand{\corot}{CoRoT\xspace}
\newcommand{\numax}{\nu_{\mathrm{max}}\xspace}
\newcommand{\lamfornacis}{$\lambda^2$ Fornacis\xspace}
\newcommand{\lamfor}{$\lambda^2$ For\xspace}
\newcommand{\lamforb}{$\lambda^2$ For b\xspace}
\newcommand{\pmodes}{\textit{p}-modes\xspace}
\newcommand{\pmode}{\textit{p}-mode\xspace}

\newcommand{\muHz}{\,\mu\mathrm{Hz}}
\newcommand{\K}{\,\mathrm{K}}
\newcommand{\Teff}{\,T_\mathrm{eff}} 
\newcommand{\FeH}{\,[\mathrm{Fe}/\mathrm{H}]}
\newcommand{\logg}{\,\log{g}}

\newcommand{\logRHK}{\log{R\mathrm{'_{HK}}}}
\newcommand{\Msun}{\,\mathrm{M}_\odot}
\newcommand{\Mstar}{\,M_{\star}}
\newcommand{\Rsun}{\,\mathrm{R}_\odot}
\newcommand{\Rstar}{\,R_{\star}}
\newcommand{\Lsun}{\,\mathrm{L}_\odot}
\newcommand{\Lstar}{\,L_{\star}}
\newcommand{\Mearth}{\mathrm{M}_\oplus}

\newcommand{\kima}{\texttt{Kima}\xspace}

\begin{document}

   \title{\tess asteroseismology of the known planet host star \lamfornacis}

   \author{M.B. Nielsen \inst{1,2,3}
          \and
          W.H. Ball \inst{1,2}
          \and
          M.R. Standing \inst{1}
          \and
          A.H.M.J. Triaud \inst{1}
          \and
          D. Buzasi\inst{4}
          \and
          L. Carboneau\inst{1,4}
          \and
          K.G. Stassun \inst{5}
          \and
          S.R. Kane \inst{6}
          \and
          W.J. Chaplin \inst{1,2}
          \and
          E.P. Bellinger\inst{2,7}
          \and
          B. Mosser\inst{8}
          \and
          I.W. Roxburgh \inst{10,1}
          \and
          Z. Çelik Orhan \inst{9}
          \and
          M. Y\i ld\i z \inst{9}
          \and
          S. Örtel \inst{9}
          \and
          M. Vrard \inst{11,19}
          \and
          A. Mazumdar \inst{20}
          \and
          P. Ranadive \inst{20}
          \and
          M. Deal \inst{11}
          \and
          G.R. Davies \inst{1,2}
          \and
          T.L. Campante \inst{11,12}
          \and
          R.A. García\inst{13,14}
          \and
          S. Mathur \inst{15, 16}
          \and
          L. Gonz\'alez-Cuesta \inst{15, 16}
          \and
          A. Serenelli \inst{17,18}
          }

   \institute{
        School of Physics and Astronomy, University of Birmingham, Birmingham B15 2TT, UK\\
              \email{m.b.nielsen.1@bham.ac.uk}
        \and
        Stellar Astrophysics Centre (SAC), Department of Physics and Astronomy, Aarhus University, Ny Munkegade 120, DK-8000 Aarhus C, Denmark
        \and    
        Center for Space Science, NYUAD Institute, New York University Abu Dhabi, PO Box 129188, Abu Dhabi, United Arab Emirates
        \and
        Department of Chemistry and Physics, Florida Gulf Coast University, 10501 FGCU Blvd., Fort Myers, FL 33965 USA
        \and 
        Department of Physics \& Astronomy, Vanderbilt University, Nashville, TN 37235, USA
        \and
        Department of Earth and Planetary Sciences, University of California, Riverside, CA 92521, USA
        \and
        School of Physics, University of New South Wales, Kensington NSW 2033, Australia
        \and
        LESIA, Observatoire de Paris, Universit\'e PSL, CNRS, Sorbonne Universit\'e, Universit\'e de Paris, 92195 Meudon, France
        \and
        Department of Astronomy and Space Sciences, Science Faculty, Ege University, 35100, Bornova, \.Izmir, Turkey.
        \and 
        Astronomy Unit, School of Physics and Astronomy, Queen Mary University of London, London E1 4NS, UK
        \and
        Instituto de Astrof\'{\i}sica e Ci\^{e}ncias do Espa\c{c}o, Universidade do Porto,  Rua das Estrelas, 4150-762 Porto, Portugal
        \and
        Departamento de F\'{\i}sica e Astronomia, Faculdade de Ci\^{e}ncias da Universidade do Porto, Rua do Campo Alegre, s/n, 4169-007 Porto, Portugal
        \and
        IRFU, CEA, Universit\'e Paris-Saclay, F-91191 Gif-sur-Yvette, France
        \and
        AIM, CEA, CNRS, Universit\'e Paris-Saclay, Universit\'e Paris Diderot, Sorbonne Paris Cit\'e, F-91191 Gif-sur-Yvette, France
        \and
        Instituto de Astrof\'{\i}sica de Canarias, La Laguna, Tenerife, Spain
        \and
        Dpto. de Astrof\'{\i}sica, Universidad de La Laguna, La Laguna, Tenerife, Spain
        \and
        Institute of Space Sciences (ICE, CSIC), Carrer de Can Magrans S/N, E-08193, Bellaterra, Spain
        \and
        Institut d'Estudis Espacials de Catalunya (IEEC), Carrer Gran Capita 2, E-08034, Barcelona, Spain
        \and
        Department of Astronomy, The Ohio State University, Columbus, OH 43210, USA
        \and
        Homi Bhabha Centre for Science Education, TIFR, V. N. Purav Marg, Mankhurd, Mumbai 400088, India
        }

   \date{Received September 15, 1996; accepted March 16, 1997}

  \abstract
   {The Transiting Exoplanet Survey Satellite (TESS) is observing bright known planet-host stars across almost the entire sky. These stars have been subject to extensive ground-based observations, providing a large number of radial velocity (RV) measurements.}
   {In this work we use the new TESS photometric observations to characterize the star $\lambda^2$ Fornacis, and following this to update the parameters of the orbiting planet $\lambda^2$ For b.}
   {We measure the p-mode oscillation frequencies in $\lambda^2$ For, and in combination with non-seismic parameters estimate the stellar fundamental properties using stellar models. Using the revised stellar properties and a time series of archival RV data from the UCLES, HIRES and HARPS instruments spanning almost 20 years, we refit the orbit of $\lambda^2$ For b and search the RV residuals for remaining variability.}
   {We find that $\lambda^2$ For has a mass of $1.16\pm0.03$M$_\odot$ and a radius of $1.63\pm0.04$R$_\odot$, with an age of $6.3\pm0.9$Gyr. This and the updated RV measurements suggest a mass of $\lambda^2$ For b of $16.8^{+1.2}_{-1.3}$M$_\oplus$, which is $\sim5$M$_\oplus$ less than literature estimates. We also detect a periodicity at 33 days in the RV measurements, which is likely due to the rotation of the host star.}
   {While previous literature estimates of the properties of $\lambda^2$ are ambiguous, the asteroseismic measurements place the star firmly at the early stage of its subgiant evolutionary phase. Typically only short time series of photometric data are available from TESS, but by using asteroseismology it is still possible to provide tight constraints on the properties of bright stars that until now have only been observed from the ground. This prompts a reexamination of archival RV data from the past few decades to update the characteristics of the planet hosting systems observed by TESS for which asteroseismology is possible.}


   \maketitle

\section{Introduction}
The Transiting Exoplanet Survey Satellite \citep[\tess,][]{Ricker2014}  observed the southern celestial hemisphere in the period from July 2018 to July 2019. The main objective of \tess is to observe short-period transiting exoplanets around bright stars. The observation strategy during the first year of operations   covered almost the entirety of the southern hemisphere, observing large swaths of the sky for short periods of time ($\approx27$ days). This is a departure from  the previous space-based photometry missions \corot \citep{Fridlund2006} and \kepler \citep{Borucki2010}, which provided photometric time series of hundreds of days or even several years for a few select fields. These time series have been a huge advantage for asteroseismology, which benefits from long observations and bright stars to make precise measurements of the oscillation modes of a star. The cohort of \tess targets extends to much brighter targets than \kepler and \corot, and so despite a lack of long baseline time series, the mission has already yielded a multitude of previously unknown variable stars, including solar-like oscillators. 

The star \lamfornacis (HD\,16417, \lamfor) was initially selected for observation in the \tess two-minute cadence mode based on its brightness (G-band magnitude of $5.59$) and high likelihood of exhibiting solar-like oscillations, as indicated by the Asteroseismic Target List \citep{Schofield2019}. It was observed for approximately two months shortly after the beginning of the \tess mission, and is one of the first planet-host stars to be observed by \tess with confirmed solar-like oscillations \citep[see also][]{Huber2019, Campante2019}. Previous studies of \lamfor have yielded a wide range of physical parameters \citep[e.g.,][]{Bond2006, Gehren1981, Bensby2014}, indicating a spectral type anywhere between G2V or G8IV. Despite the relatively short amount of time that this star was observed by TESS, the photometric variability shows an unambiguous power excess at a frequency of $\approx1280\muHz$, caused by solar-like acoustic ($p$-mode) oscillations. Stars with outer convective zones like the Sun and \lamfor oscillate with regularly spaced overtones of radial and non-radial modes with angular degree, $l$. These modes propagate through the interior of the star, and therefore place tight constraints on its physical properties  \citep[e.g.,][]{Garcia2019}.  The oscillation power of \lamfor peaks at around $1280\muHz$, which alone places the star firmly in the subgiant regime. However, going a step further and measuring the individual mode frequencies has been shown to yield estimates of the mass and radius at a precision of a few percent, and the stellar age at $\approx10\%$ \citep[e.g.,][]{Brown1994, Lebreton2014, Angelou2017, Bellinger2019b}. This has implications for estimates of the characteristics of any potential orbiting planets, in particular with respect to the mass of the planets, but also in terms of the dynamical history of the system itself. 

\citet{Otoole2009} discovered a roughly Neptune-mass planet in a $17.25$-day orbit around \lamfor. The detection was made using radial velocity (RV) measurements from the UCLES spectrograph at the Anglo-Australian Telescope (AAT), and the HIRES spectrograph at the Keck telescope. Since then an extensive set of HARPS \citep{Mayor2003} data from the ESO La Silla 3.6m telescope has become publicly available (see Table \ref{tab:lit}). Here we combine the original observations by \citet{Otoole2009} and the HARPS measurements to construct an almost 20-year set of RV data. This, combined with the updated estimates of the stellar mass from asteroseismology, allows us to better constrain the planet and its orbital parameters. Furthermore, the extensive RV  data set and the high quality of the HARPS measurements presents an opportunity to investigate variability on timescales other than the orbit of the known planet.

The \tess time series reduction is presented in Section \ref{sec:TS}. The modeling process of \lamfor is described in Section \ref{sec:modeling}, including the power spectrum and spectral energy density SED fitting processes, which yields the seismic and non-seismic constraints, respectively. In Section \ref{sec:lam4b} we discuss the methods used to improve the estimates of the planet and orbital characteristics of the known planet and the methods used to investigate additional periodicity in the RV measurements. 

\section{Time series preparation}
\label{sec:TS}
\lamfor was observed by \tess in Sectors $3$ and $4$, for a total of approximately $2$ months. The photometric time series are available in a pre-reduced version from the Science Processing Operations Center (SPOC) pipeline \citep{Jenkins2016}, and in the form of pixel-level data. The SPOC pipeline removes artifacts and carries out traditional CCD data reduction activities, such as bias correction and flat-fielding, prior to performing aperture photometry. Aperture sizes are computed using the algorithm originally developed for \kepler postage stamps, which makes use of stellar parameters from the TESS Input Catalog \citep{Stassun2018c} together with models of the detector and spacecraft attitude control system. Following extraction, the presearch data conditioning (PDC) algorithm removes instrumental signatures due to changes in pointing and focus, and performs corrections due to stellar crowding and aperture overfilling by the target star. 

We attempted to improve on the SPOC result by manually reducing the pixel-level data. This was done using a procedure that has  previously produced better signal-to-noise ratio (S/N) oscillation spectra for asteroseismic analyses of bright stars compared to the standard SPOC pipeline (Metcalfe et al. in prep). The approach broadly mirrors that used successfully on K2 data in the past \citep[see, e.g.,][]{Buzasi2015}, and involved defining a custom photometric aperture pixel mask. In the case of \lamfor this procedure did not reduce the noise level around the \pmode envelope, but for completeness we briefly summarize the method and the result in Appendix~\ref{app:manual}. In the following sections we only use the SPOC PDC time series.

\section{Modeling \lamfor}
\label{sec:modeling}
To estimate the fundamental properties of \lamfor we used both seismic and non-seismic constraints. The seismic constraints came from measuring the individual oscillation frequencies observed in the power spectrum of the SPOC time series, also known as peakbagging. The non-seismic constraints were derived from SED fitting. These constraints were then used by several independent modeling teams to provide estimates of the mass, radius, and age of \lamfor. These steps are detailed in the following section. 

\subsection{Seismic constraints}
\begin{figure*}
    \centering
    \includegraphics[width = 2\columnwidth]{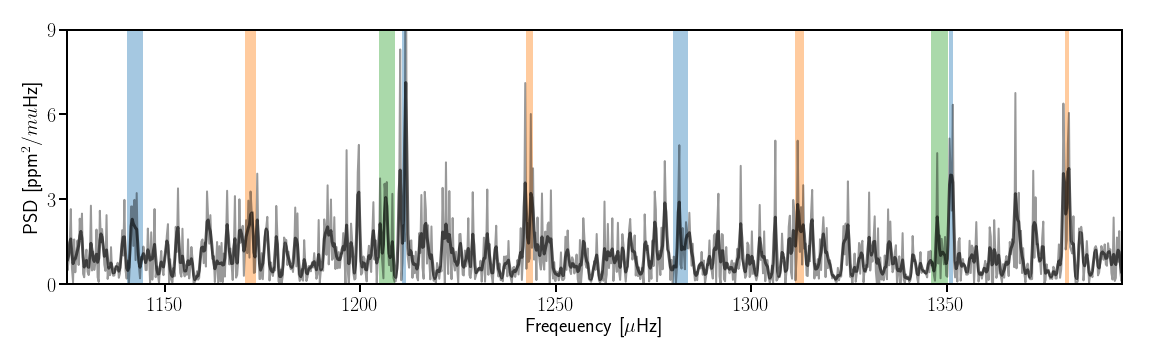}
    \caption{Power density spectrum (gray) and the smoothed spectrum (black) around the \pmode oscillation frequencies. The $68\%$ confidence interval of the fit mode frequencies are shown as the vertical shaded regions. Colors denote the angular degree, $l$, where blue is $l=0$, orange is $l=1$, and green is $l=2$.}
    \label{fig:psd}
\end{figure*}

\begin{figure}
    \centering
    \includegraphics[width = \columnwidth]{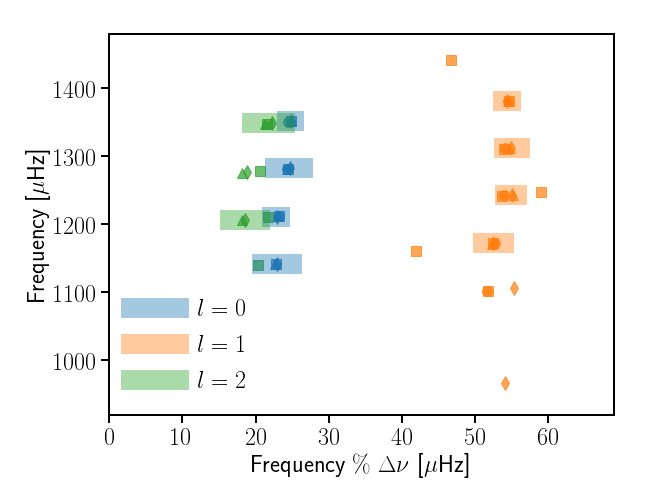}
    \caption{\'Echelle diagram showing oscillation frequencies modulo the large separation ($\Delta\nu=69.0\muHz$) supplied by each team: MV (circles), IWR (squares), MBN (diamonds), and WJC (triangles). The colors represent the angular degrees, $l$, that were considered. The shaded regions represent $68\%$ confidence interval of the frequencies in the final fit. Several mixed modes (diverging modes along the $l=1$ ridge) were suggested, but could not be verified by the other teams, and so were not included in the final fit.}
    \label{fig:echelle}
\end{figure}

To measure the oscillation frequencies of \lamfor we used the SPOC flux time series as described above. Because of the low S/N of the oscillations, several independent peakbaggers\footnote{Peakbagging team members: MBN, IWR, MV, BM, WJC} were tasked with finding and fitting the oscillation peaks. Initial guesses for the frequencies were found using the universal pattern approach \citep[see][]{Mosser2013} and by-eye inspection. The final choice of which modes to fit was based on the subset that all teams agreed on within their respective uncertainties. Each mode alone only has a $\sim 95\%$ probability of not being due to background noise, whereas this manual approach also incorporates knowledge of the repeating pattern of the \pmodes. In a low S/N case like \lamfor, this repeating pattern helps identify the initial mode frequencies. The mode frequencies range from $1142\muHz$ to $1380\muHz$, with a peak of oscillation power at $\numax \approx 1280\muHz$ and a separation of consecutive overtones of the same angular degree (large separation), $\Delta\nu \approx 69\muHz$. 

The final list of frequencies was then fit using a Markov chain Monte Carlo (MCMC) approach, the result of which was used in the remaining analysis of \lamfor. The model used to fit the power spectrum mirrors that used in previous peakbagging efforts such as \citet{Handberg2011} or \citet{Lund2017}. The mode frequencies and heights were treated as independent variables for each mode, and the mode widths were assumed to follow the relation by \citet{Appourchaux2014}. The fit assumed a single rotational splitting and inclination axis for all oscillation modes. The background noise levels from granulation, activity, and uncorrected instrumental effects were fit concurrently with the modes, using two Harvey-like profiles \citep{Harvey1988, Kallinger2014}. We applied uniform priors on the mode frequencies, background timescales, rotational splitting, inclination angle, and the location parameters of the mode width relation \citep[see][]{Appourchaux2014}. For the remaining model parameters we used log-uniform priors. 

The posterior distribution of the model parameters was mapped using an MCMC sampler\footnote{\texttt{emcee}: \citet{Foreman-Mackey2013}}. For each model parameter the median of the marginalized posterior distribution was taken as the  best-fit solution, and the $16$th and $84$th percentile interval as a measure of the parameter uncertainties. 

The power density spectrum and the resulting fit frequencies are shown in Fig. \ref{fig:psd}, and the \'echelle diagram in Fig. \ref{fig:echelle}. Modes of angular degree $l=3$ were not considered by any of the teams as these are typically very low amplitude, and thus require exceptional S/N to be observed. One team suggested the possible presence of mixed $l=1$, but this could not be verified by the other teams and so they were not included in the final fit. The final list of fit frequencies is presented in Table \ref{tab:freqs}. 

The fit to the power spectrum was unable to constrain the rotational splitting of the modes to less than $\approx2.1\muHz$, and thus the stellar rotation rate. Similarly, the seismic data were unable to constrain the stellar inclination angle. The marginalized posterior distributions of the inclination angle and rotational splitting are shown in Fig. \ref{fig:banana}.

\begin{figure}
    \centering
    \includegraphics[angle=90, width = \columnwidth, trim=20 20 30 40]{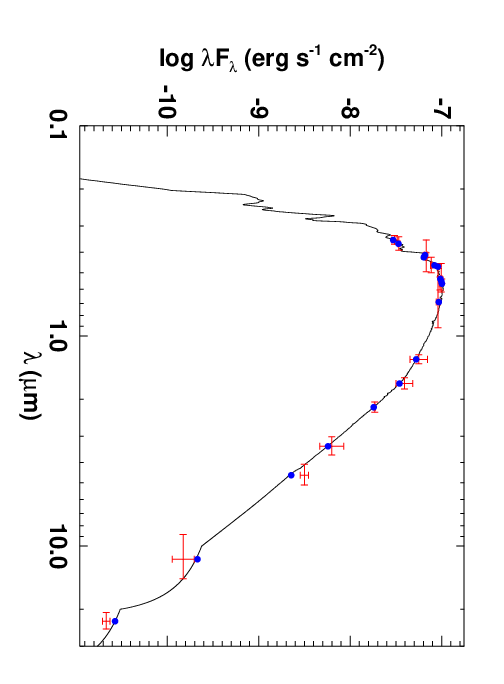}
    \caption{Spectral energy distribution of \lamfor. Red symbols represent the observed photometric measurements, where the horizontal bars represent the effective width of the passband. The photometric measurements are $B_T V_T$ magnitudes from {\it Tycho-2}, the $BVgri$ magnitudes from APASS, the $JHK_S$ magnitudes from {\it 2MASS}, the W1--W4 magnitudes from {\it WISE}, and the $G$ magnitude from {\it Gaia}. Blue symbols are the model fluxes from the best-fit Kurucz atmosphere model (black). 
    \label{fig:sed}}
\end{figure}

\begin{table*}
\centering
\caption{Stellar model settings for the different teams.  A single entry is used where all three teams used the same input physics.}
\begin{tabular}{cccc}
\hline
Team & Birmingham & Mumbai & Porto \\
\hline
Models & MESA${}^a$ (r10398) & MESA${}^a$ (r10398) & GARSTEC${}^b$ \\
Oscillations & GYRE & GYRE & ADIPLS \\
High-$T$ opacities & \multicolumn{3}{c}{--- OPAL \citep{Iglesias1993,Iglesias1996} ---} \\
Low-$T$ opacities & \multicolumn{3}{c}{--- \citet{Ferguson2005} ---} \\
Gravitational settling & \multicolumn{3}{c}{--- \citet{Thoul1994} ---} \\
EoS & MESA/OPAL & MESA/OPAL & FreeEOS \\
Solar mixture & GN93 & GS98 & GN93 \\
Helium enrichment law ($Y=\ldots$) & $1.28855Z+0.248$ & $2Z+0.24$ & $1.1843Z+0.2485$ \\
Nuclear reactions & NACRE & NACRE & Solar Fusion II \\
Atmosphere & \citet{Mosumgaard2018} & \citet{Krishna1966} & Eddington \\
$\alpha_\mathrm{MLT}$ & $1.037$* & $1.81$, $1.91$, $2.01$ & $1.811$ \\
Surface correction & BG14-1 & BG14-1 & BG14-2\\
Overshooting & None & $0$ \& $0.016$ & $0.02$\\
\hline
\end{tabular}
\\
${}^a$ \citet{Paxton2011,Paxton2013,Paxton2015}\quad 
${}^b$ \citet{Weiss2008}\quad
\citet{Ball2014}\quad
\citet{Adelberger2011}\quad
\citet{Rogers2002}\quad
\citep{Irwin2012}\quad
\citet{Grevesse1993}\quad
\citet{Townsend2013,Townsend2018}\quad
\citet{Herwig2000}
\label{tab:physics}
\end{table*}

\subsection{Stellar modeling}
\label{sec:stellarmodeling}
Three teams, identified by their principal locations, independently fit stellar models to the seismic and non-seismic data for \lamfor. The teams independently chose stellar evolution codes, stellar pulsation codes, non-seismic observables, and fitting methods. The main choices of input physics are summarized in Table \ref{tab:physics} and the best-fitting parameters of the models, with uncertainties, are listed in Table \ref{tab:modelvalues}. For the frequencies derived from seismology, all the teams used either the one- or two-term surface correction by \citet{Ball2014}.  We did not enforce a line-of-sight velocity correction as this was negligible for \lamfor \citep{Davies2014, Soubiran2018}. We describe below more complicated details of the stellar models, and how each team fit their models to the data. Our final estimates of the stellar properties are precise to $3$ \%\ in mass, $2.7$ \%\ in radius, and $14$ \% in age. 

\subsubsection{Birmingham}
\label{sssec:bham}

The Birmingham team used Modules for Experiments in Stellar Astrophysics \citep[MESA, r10398;][]{Paxton2011,Paxton2013,Paxton2015} with the atmosphere models and calibrated mixing-length (MLT) parameters from \citet{Trampedach2014a,Trampedach2014b} as implemented in \citet{Mosumgaard2018}. The mixing-length parameter in Table \ref{tab:physics} is the calibrated correction factor that accommodates slight differences between MESA's input physics and  mixing-length model and that of the simulations by \citet{Trampedach2014a,Trampedach2014b}, rather than the mixing-length parameter $\alpha_\mathrm{MLT}$. The free parameters in the fit are the stellar mass $M$, the initial metallicity $\FeH_i$, and the age $t$. 

The free parameters were optimized by first building a crude grid based on scaling relations, then optimizing the best model from that grid using a combination of a downhill simplex \citep[i.e., Nelder--Mead method,][]{Nelder1965} and random resampling within error ellipses around the best-fitting parameters when the simplex stagnated. Uncertainties were estimated by the same procedure as used by \citet{Ball2017}.

The objective function for the optimization was the unweighted total $\chi^2$ of both the seismic and non-seismic data, using observed non-seismic values of $\Teff$, $\FeH$, and $\Lstar/\Lsun$. The $\Teff=5841\pm60\K$ and $\FeH=0.13\pm0.06$ values were taken from \citet{DelgadoMena2017}, with uncertainties increased to those used for most of the stars in \citet{Lund2017}. The luminosity was derived from SED fitting following the procedures described in \citet{Stassun2016}, \citet{Stassun2017}, and \citet{Stassun2018a}. The available photometry spans the wavelength range 0.35--22~$\mu$m (see Figure~\ref{fig:sed}). We fit the SED using Kurucz stellar atmosphere models \citep{Kurucz2013}, with the priors on effective temperature $T_{\rm eff}$, surface gravity $\log g$, and metallicity [Fe/H] from the spectroscopically determined values. The remaining free parameter in the SED fit is the extinction ($A_V$), which we restricted to the maximum line-of-sight value from the dust maps of \citet{Schlegel1998}. The best-fit extinction is $A_V = 0.04 \pm 0.04$. Integrating the  model SED (which is unreddened) gives the bolometric flux at Earth of $F_{\rm bol} = 1.36 \pm 0.05 \times 10^{-7}$ erg~s~cm$^{-2}$. Together with the \textit{Gaia} parallax of $39.3512\pm0.0534\,\mathrm{mas,}$ this yielded a constraint for the luminosity of $\log \Lstar/\Lsun=0.436\pm0.015$.

\subsubsection{Mumbai}
The Mumbai team computed a grid of stellar models also using MESA (r10398).  The grid spans masses from $1.10$ to $1.38\Msun$ in steps of $0.01\Msun$, initial metallicities $\FeH_i$ from $-0.02$ to $0.36$ in steps of $0.02$, and mixing-length parameters $\alpha_\mathrm{MLT}$ of $1.81$, $1.91$, and $2.01$.  Gravitational settling, which is otherwise included in the stellar models, is disabled for models with $\Mstar>1.3\Msun$, but the best-fitting models are less massive and unaffected by this choice.
The grid uses two values for the length scale of core convective overshooting, using the exponentially decaying formulation by \citet{Herwig2000}: $f_\mathrm{ov}=0$ (i.e., no overshooting) and $0.016$.

The goodness of fit was evaluated through a total misfit defined by
\begin{equation}
  \chi^2_\mathrm{Mum} = (\chi^2_{\Teff} + \chi^2_{\log g} + \chi^2_{\FeH} + \chi^2_\nu )
,\end{equation}
where for $x=\Teff$, $\log g$, or $\FeH$:
\begin{equation}
  \chi^2_x=\left(\frac{x_\mathrm{model} - x_\mathrm{obs}}{\sigma_x}\right)^2
\end{equation}
and
\begin{equation}
  \chi^2_\nu = \frac{1}{N}\sum_i\left(\frac{\nu_{i,\mathrm{model}}- \nu_{i,\mathrm{obs}}}{\sigma_{\nu,i}}\right)^2
.\end{equation}
The observed values of the non-seismic data were $\Teff=5790\pm150\K$, $\log g=4.11\pm0.06$, and $\FeH=0.09\pm0.11$.
The reported parameters are likelihood-weighted averages and standard deviations of the likelihood evaluated for each model in the grid, where the unnormalized likelihood is
\begin{equation}
  \mathcal{L}_\mathrm{Mum}=\exp\left(-\frac{1}{2}\,\chi^2_\mathrm{Mum}\right)
.\end{equation}

\subsubsection{Porto}
The Porto team used Asteroseismic Inference on a Massive Scale \citep[AIMS:][]{Lund2018,Rendle2019} to optimize a grid of stellar models computed with GARSTEC \citep{Weiss2008}. The observed non-seismic values were taken to be $\Teff=5792.5\pm143.5\K$, metallicity $\FeH= 0.09\pm0.10$, and luminosity $\Lstar/\Lsun=2.71\pm0.10$. The luminosity was determined from the relation by \citet{Pijpers2003} using the extinction correction derived from the SED fit described in Section \ref{sssec:bham}. The masses $M$ in the grid ranged from $0.7$ to $1.6\Msun$ in steps of $0.01\Msun$ and the initial metallicity $\FeH_i$ ranged from $-0.95$ to $0.6$ in steps of $0.05$.  The models included extra mixing below the convective envelope according to the prescription by \citet{Vandenberg2012}.  The efficiency of microscopic diffusion is smoothly decreased to zero from $1.25$ to $1.35\Msun$ though, again, the best-fitting models are all significantly below $1.25\Msun$ and therefore not affected by this choice.  In addition, a geometric limit is applied for small convective regions, as described in \citet{Magic2010}.

The goodness-of-fit function was the unweighted total $\chi^2$ of the seismic and non-seismic data, as used by the Birmingham team (Sec.~\ref{sssec:bham}).

\subsection{Adopted fundamental stellar parameters}
Table \ref{tab:modelvalues} includes parameters averaged across the three stellar model fits. Specifically, we took the average and $1\sigma$ percentile ranges from the evenly weighted combination of the three fits.  

\begin{figure}
    \centering
    \includegraphics[width = \columnwidth]{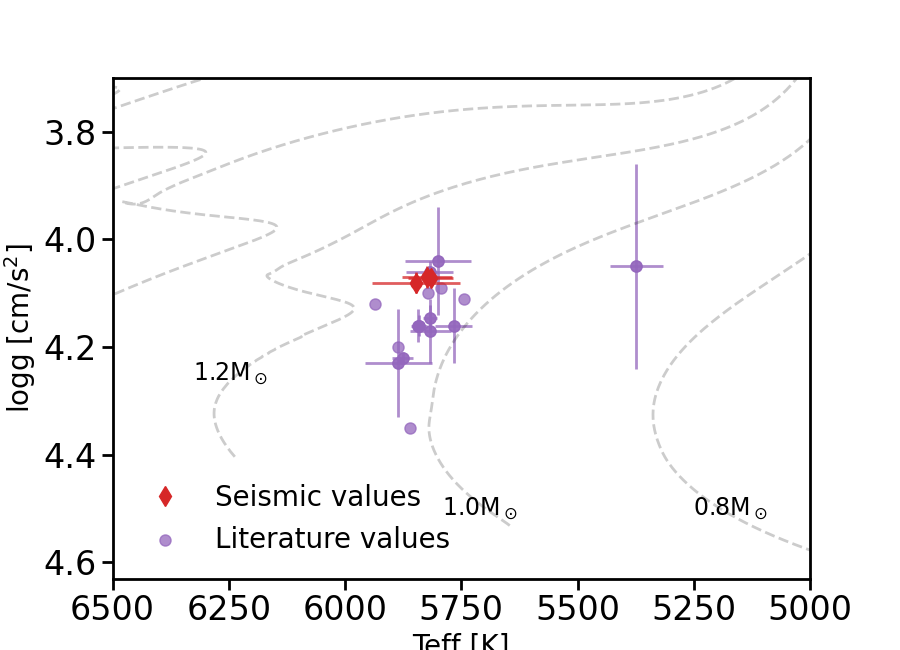}
    \includegraphics[width = \columnwidth]{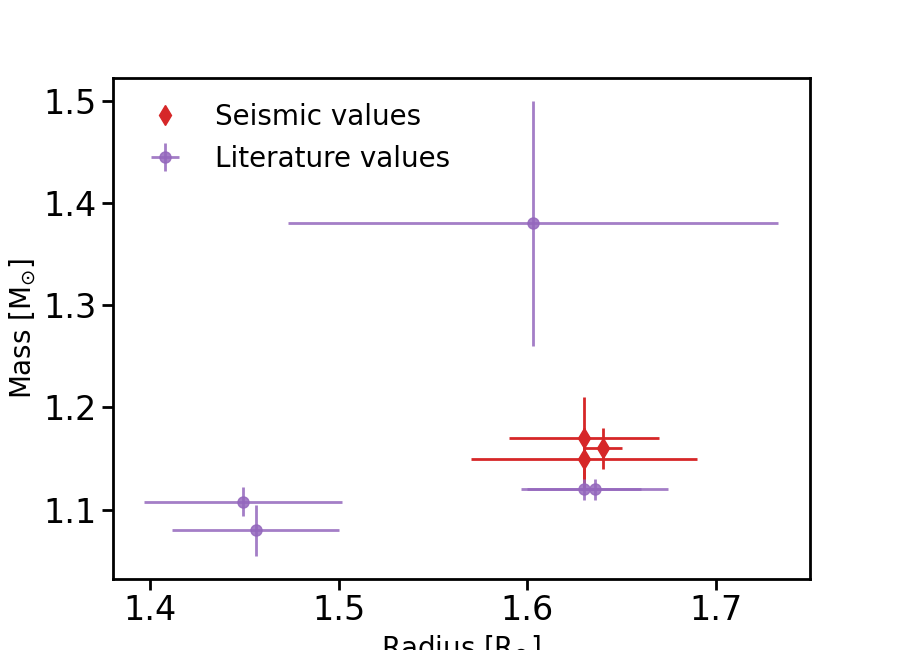}
    \caption{Top: Kiel diagram showing the literature values (purple) and seismic values (red; see Table \ref{tab:modelvalues}) of \lamfor. Dashed lines indicate evolutionary tracks spanning a mass range of $0.8-2\Msun$, in increments of $0.2\Msun$. Literature values are presented in Table \ref{tab:lit}. Bottom: Masses and radii of the literature sources with a combination of $\logg$ with either mass or radius in comparison to the seismic estimates.}
    \label{fig:comparison}
\end{figure}

Figure \ref{fig:comparison} shows the $\logg$ and $\Teff$ values estimated by each team in relation to the literature values (see also Table \ref{tab:lit}). Among the more recent literature sources the estimates of $\logg$ span a considerable range of $\approx0.2$ dex, and a subset of sources reported a mass ranging from $1.08^{+0.03}_{-0.02}\Msun$ to $1.38\pm0.12\Msun$, and corresponding radius between $1.45\pm0.05\Rsun$ and $1.61\pm0.13\Rsun$ \citep[see bottom frame of Fig. \ref{fig:comparison},][]{Valenti2005, Ramirez2014}.

The asteroseismic measurements allow us to obtain robust estimates of the surface gravity, and subsequently of the mass and radius. Despite the inclusion of different model physics and approaches taken by the modeling teams in this work, the resulting estimates all fall within a few percentage points of each other in both mass and radius, with average values of $\Mstar=1.16\pm0.03\Msun$ and $\Rstar=1.63\pm0.04\Rsun$. There is therefore agreement between the asteroseismic modeling results that \lamfor is at the early stages of the subgiant evolutionary phase, whereas previous estimates were unable to confirm this unambiguously. 

The literature estimates of the age \citep{Valenti2005, Tsantaki2013, Bonfanti2016} fall within $1-2\sigma$ of the asteroseismic estimates of $6.3\pm0.9$~Gyr, with the extremes at $4.3\pm0.8$~Gyr \citep{daSilva2006} and $7.6\pm0.7$ \citep{Nordstroem2004}.

\begin{table*}
    \centering
    \caption{Model parameters for \lamfor using seismic and non-seismic constraints.}
    \begin{tabular}{cccccc}
      \hline
      Team & Birmingham & Mumbai & Porto & Adopted \\
      \hline
      Mass [$\Msun$]             & $1.16\pm0.02$   & $1.17\pm0.04$ & $1.15\pm0.03$ & $1.16\pm0.03$\\
      Radius [$\Rsun$]           & $1.64\pm0.01$   & $1.63\pm0.04$ & $1.63\pm0.06$ &$1.63\pm0.04$\\
      Age [Gyr]                  & $6.4\pm0.5$     & $5.6\pm1.0$   & $6.7\pm0.8$ &$6.3\pm0.9$\\
      $\Teff$ [K]                & $5816\pm47$     & $5847\pm80$   & $5824\pm102$ & $5829\pm80$\\
      $\log{g}$ [cm/s$^2$]       & $4.07\pm0.01$   & $4.08\pm0.02$ & $4.07\pm0.04$ & $4.08\pm0.03$\\
      $[\mathrm{Fe}/\mathrm{H}]$ & $0.12\pm0.07$   & $0.09\pm0.06$ & $0.11\pm0.10$ & $0.10\pm0.08$\\
      \hline
    \end{tabular}
    \label{tab:modelvalues}
\end{table*}

\section{Radial velocity analysis of the \lamfor system}
\label{sec:lam4b}
Since the discovery of \lamforb, a much larger sample of RV measurements has become available from the HARPS spectrometer at the ESO La Silla 3.6m telescope. This presents an opportunity to update the orbital parameters of the planet based on this new data, and by using the new estimates of the stellar mass from asteroseismology. The HARPS data were downloaded via the ESO Science Portal\footnote{The HARPS data set was collected thanks to several observing programs listed in Appendix \ref{tab:progID}, and were obtained from: \myurl}. We combined these data with the AAT and Keck data as presented by \citet{Otoole2009} in Table 2 of that publication. 

\subsection{Updated characteristics \lamforb}
We used \kima \citep{kima} to analyze the combined data set. \kima fits a sum of Keplerian curves to RV data corresponding to one or more potential planets. It uses a diffusive nested sampling algorithm \citep{Brewer2009} to sample the posterior distribution of parameters explaining the data, where in this case the number of planets $N_p$ was left as a free parameter. This allowed us to use \kima to estimate the fully marginalized Bayesian evidence of the parameter space, which was used to determine the likelihood of any number of planets that may be present and detectable in the data. 
 
\begin{figure}
    \centering
    \includegraphics[width = \columnwidth]{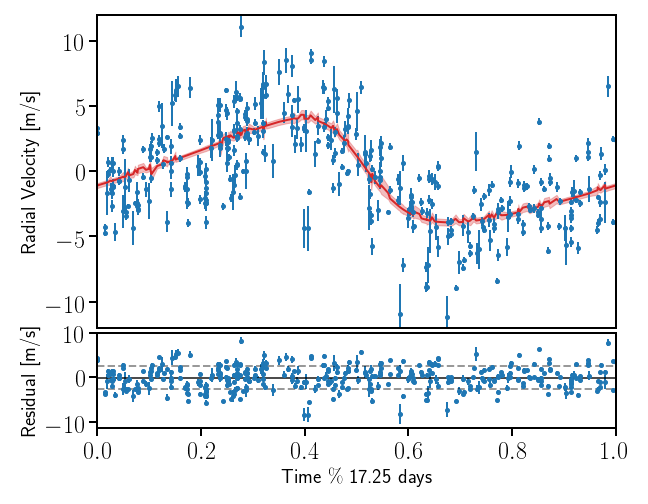}
    \caption{Top: Observational RV data (blue) of \lamfor, from AAT, Keck, and HARPS, phase-folded at a period of $17.25$ days. The phase-folded best-fit model is shown in red, with the model uncertainties in shaded red. Bottom: Residual RV after subtracting the best-fit RV model. The dashed lines indicate the standard deviation, $\sigma = 2.64$~m/s, of the residual.}
    \label{fig:rvmodel}
\end{figure}

 For the analysis of the RV data from \lamfor, $N_p$ was set as a free parameter with an upper limit of $N_p=5$. Once samples of the fit posterior were obtained from \kima any proposed crossing orbits were removed a posteriori. The resulting posterior consists of a wide parameter space with a number of overdensities corresponding to regions of high likelihood for each of the parameters, such as orbital period $P$, semi-amplitude $K$, and eccentricity $e$. We identified these regions with the clustering algorithm \texttt{HDBSCAN} \citep{McInnes2017}. \texttt{HDBSCAN} identified a cluster corresponding to the orbital period of \lamforb. We extracted these samples from the posterior and used them to approximate the posterior probability density of the planetary orbital and physical parameters. As a result we provide the median of the distribution of each parameter (see figures \ref{fig:np} and \ref{fig:planetcorner}), and provide uncertainties estimated from the 16th and 84th percentiles, which are shown in Table \ref{table:HD16417b_table}. The best-fit model is shown in Fig. \ref{fig:rvmodel}, along with the residual RV signal, which has a standard deviation of $\sigma=2.64$~m/s. 
 
 To verify the \kima results, we also fit the RV data using the \texttt{exoplanet} package \citep{exoplanet}, which, like \kima, also fits Keplerian orbits, but includes the signal from stellar granulation noise as a Gaussian process in the RV model. The \texttt{exoplanet} package uses the \texttt{celerite} library \citep{celerite} to model any variability in the RV signal that can be represented as a stationary Gaussian process. The choice of kernel for the Gaussian process was set by the quality factor which we here chose to be $Q=1/2$ to represent a stochastically excited damped harmonic oscillator with a characteristic timescale $w_0$ and amplitude $S_0$. We applied a normal prior on $w_0$ based on the modeling of the TESS power spectrum presented in Section \ref{sec:modeling} as the granulation timescale is expected to be identical in both radial velocity and photometric variability. The granulation power in the TESS intensity spectrum is not easily converted to a radial velocity signal as seen from multiple different instruments, we therefore use a weakly-informative log-normal prior on $S_0$. We found that the \kima and \texttt{exoplanet} results were consistent within $1\sigma$. 
  
\begin{table}
\centering
\onehalfspacing
\caption{Best-fit orbital parameters of \lamforb}
\begin{tabular}{ccc} 
 \hline
 Parameter & Discovery paper & This work \\ [0.5ex] 
 \hline
 $P$ [days] & $17.24 \pm 0.01$ & $17.251^{+0.002}_{-0.003} $ \\ 
 $K$ [m/s] & $5.0 \pm 0.4$ & $4.0\pm0.3 $ \\
 $e$ & $0.20 \pm 0.09$ & $0.35^{+0.05}_{-0.05} $ \\
 $m_p \sin{i}$ [$\Mearth$] & $22.1 \pm 2.0$ & $16.8^{+1.2}_{-1.3} $ \\ [1ex]
 \hline
\end{tabular}
\tablefoot{Orbital parameters of \lamforb, where $P$ is the orbital period, $K$ is the RV semi-amplitude, $e$ is the orbital eccentricity, and $m_p \sin{i}$ is the estimated lower limit of the planet mass. The middle column shows the values found by \citet{Otoole2009}, the right column shows the best-fit parameters from \kima found in this work.}
\label{table:HD16417b_table}
\end{table}

\subsection{Additional radial velocity variability}
\begin{figure}
    \centering
    \includegraphics[width=\columnwidth]{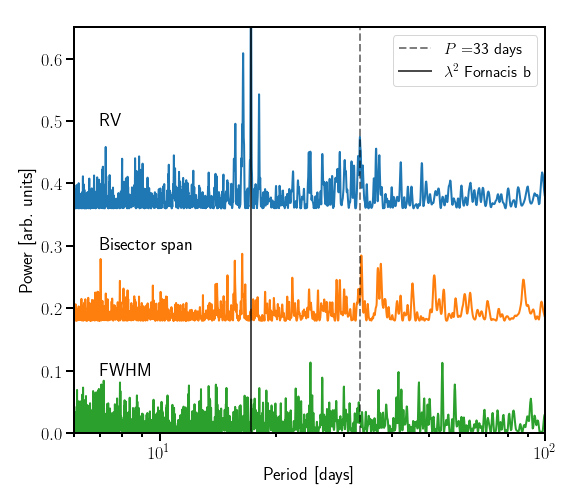}
    \caption{Periodograms of the HARPS RV data (blue), the spectral line bisector span (orange), and the cross-correlation function full-width at half maximum (FWHM, green). The FWHM and bisector span were only available for the HARPS data, and so the AAT and Keck data are not included in the power spectra shown here. The full vertical line shows the period of \lamforb, and the dashed line is the secondary $33$-day periodicity. The comb of peaks around the orbital period of \lamforb are caused by the observational window function, as is the case for many of the peaks around the $33$-day periodicity.}
    \label{fig:rvperiod}
\end{figure}

\citet{Otoole2009} suggested the presence of a periodicity at $\approx298$ days, which they ultimately did not attribute to the presence of an additional planet. The posterior distribution of the fit parameters obtained from \kima indeed shows a periodicity at $\approx300$ days, but the Bayesian evidence does not support the added model complexity that comes from adding a planet at or near this orbital period. This was quantified by calculating the Bayes factor \citep[the ratio of evidence weighted probabilities,][]{Kass1995} for an increase in the number of planets from $N_p = 1$, i.e., assuming only \lamforb exists, to $N_p = 2$ (see Fig. \ref{fig:np}). This yielded a Bayes factor of $\approx 1.67$,  which is ``not worth more than a bare mention'' \citep{Kass1995}. This was found to be the case when testing both just the AAT and Keck data set, and with the added HARPS data. 

However, using the combined data sets highlights a period at $\approx 33$ days. Figure \ref{fig:rvperiod} shows the periodogram of the RV measurements, where the signal due to \lamforb is visible at $P=17.25$ days, with surrounding aliases caused by the observational window function. The $33$-day periodicity shows aliasing in the RV, but also appears in the bisector span, which in addition shows harmonic peaks at a period of $\approx 16.5$ days. This signal was not discussed by \citet{Otoole2009}, and so to investigate this periodicity further we established two scenarios: first, that it is due to another planet in a wider orbit than the known planet or second, that it is due to variability induced by magnetic activity on the stellar surface.

\subsubsection{Scenario 1: An additional planet}
In the first scenario we consider that the $33$-day signal is due to an additional planet in the \lamfor system. From Fig. \ref{fig:rvperiod} a bisector span variation is apparent at this period, which is a strong indication that the signal is not a planet \citep[see, e.g.,][]{Queloz2001}. In addition, we used \kima to evaluate the possibility of a planet in such an orbit, but the Bayesian evidence for the additional planet remains small. This in itself would suggest that the existence of a second planet is unlikely but as a final belts-and-braces measure we tested the dynamical stability of such a planet to investigate whether it could survive on timescales comparable to the $\approx6.3$~Gyr lifetime of the \lamfor system.

The dynamical simulations were performed using the \texttt{REBOUND}  package\footnote{\href{https://rebound.readthedocs.io/en/latest/}{rebound.readthedosc.io}}, described in detail by \citet{rein2012}, and using the WHFast integrator \citep{Rein2015}. \texttt{REBOUND} computes the Mean Exponential Growth factor of Nearby Orbits \citep[MEGNO,][]{Maffione2011}, which is a chaos indicator on a logarithmic scale that quantifies the divergence of a test particle placed in relation to known orbiting planet, in this case \lamforb. These techniques have been applied to a range of exoplanetary systems \citep{gozdziewski2001a,gozdziewski2002,satyal2013,satyal2014,Triaud2017,kane2019c}.

\begin{figure}
    \centering
    \includegraphics[width=\columnwidth]{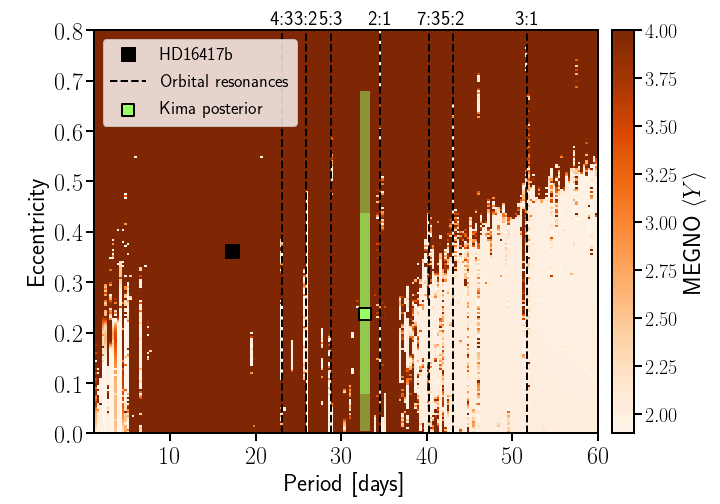}
    \caption{Stability map (MEGNO)  for a particle at different periods and inclinations, in the presence of the known planet \lamforb. The color bar indicates the linear scale of the MEGNO statistic, for which darker colors represent a higher degree of orbital divergence (chaos) on timescales of $10^5$ orbits of \lamforb. Lighter shaded regions denote stable orbits. The upper range of the color bar has been truncated at a MEGNO value of $4$ for clarity. The marginalized posterior distribution of the $33$-day orbit is shown in green, with the median indicated by the square symbol. Orbits in resonance with \lamforb are indicated by vertical dashed lines. Additional resonances are not marked for clarity. The solid black square denotes the period and eccentricity of \lamforb. }
    \label{fig:megno}
\end{figure}

Figure~\ref{fig:megno} shows the results of our dynamical simulation, where the MEGNO values are presented as a function of eccentricity and orbital period. 
MEGNO values $\lesssim2$ indicate very likely stable orbits, while values $\gtrsim2$ are either approaching instability (chaos), or for MEGNO $\gg 2$ have already diverged at the end of the simulation. The simulations were run for $10^5$ orbits of \lamforb, equivalent to $\approx$4700 years, where the configuration seen in Fig. \ref{fig:megno} was reached after just a few hundred orbits. Figure \ref{fig:megno} also shows the marginalized posterior distribution of the eccentricity and period obtained from \kima, for the $33$-day periodicity. The range of periods is narrower than the symbol size (see Table \ref{table:HD16417b_table}), but the eccentricity spans a wide range. None of the orbits within the period range have MEGNO values $\lesssim2$, indicating that any orbit at this period would be unstable. A number of very narrow stable regions appear at multiple different periods. These are all likely due to resonances with \lamforb. However, none fall near the $33$-day periodicity, excluding the possibility that a planet in this orbit could be stable due to a resonance. 

\subsubsection{Scenario 2: Stellar activity}
\citet{Otoole2009} did not discus  the $33$-day RV variability; however,  they estimated a rotation period of the star of $22-33$ days, based on the measured $\logRHK=-5.08$ and the age-activity relation by \citet{Wright2004}. Estimates of the projected rotational velocity $v\sin{i}$ in the literature fall in the range $2.1-2.5~\mathrm{km/s}$ \citep{Nordstroem2004, Valenti2005, AmmlerVonEiff2012}, which is consistent with a rotation period of $33$ days when assuming a stellar radius of $1.63~\Rsun$ and an inclined rotation axis on the order of $i\approx50^{\circ}$ relative to the line of sight to the observer. 

While the asteroseismic fit included the rotation rate and inclination of the rotation axis of the star as free variables, the marginalized posteriors for these parameters (see Fig. \ref{fig:banana}) could only constrain the rotation rate to $\lesssim2.1\muHz$. This is likely in part due to the low frequency resolution of the power spectrum ($0.2~\muHz$) compared to the expected slow rotation rate of the star ($\approx0.36~\muHz)$. Measuring rotation rates from the oscillation modes may also be further hampered by a very low angle of inclination of the rotation axis relative to the observer. 

No signatures of star spots are visible in the \tess photometry, either in the SPOC or manually reduced light curves (see Fig. \ref{fig:LC_comparison}), which might also be expected for a star with a low inclination angle and low activity level.

If this signal is indeed due to rotation, a similar period is expected in the spectral line bisector, and in some cases in the full width at half maximum (FWHM) of the cross-correlation function used to measure the RV \citep[e.g.,][]{Dumusque2012, GomesdaSilva2012}. In the case of \lamfor  the power spectra of the HARPS RV data and the bisector span (see Fig. \ref{fig:rvperiod}) show a power excess at around $33$ days, but the FWHM spectrum does not show any clear peaks at this period. Furthermore, no significant correlation was found between the residual RV, after removing the signal from \lamforb, and the bisector span where a negative correlation is in some cases an indication that the bisector span variation is caused by stellar activity \citep{Huelamo2008, Queloz2009}.  

Using the gyrochronology relation by \citet{Barnes2003}, with the asteroseismic ages from Section \ref{sec:modeling} and a $B-V=0.66$ \citep{Ducati2002} as input, we find that the rotation period of the star is likely between $P_{\mathrm{rot}}=27-31$ days. While correcting the $B-V$ estimate for interstellar reddening decreases this estimated period range, we note that the \citet{Barnes2003} relation is only calibrated for young main-sequence stars and does not account for structural evolution that occurs after stars  leave the main sequence. This is expected to increase the surface rotation period as the radius of the stellar envelope increases after the main sequence \citep[see, e.g., Fig. 3 in][]{VanSaders2016}.

\section{Discussion and  conclusions}
We used the recent release of TESS photometric data to perform an asteroseismic analysis of the star \lamfor. This allowed us to place tighter constraints on the stellar parameters than has previously been possible. We measured individual oscillation mode frequencies of the star centered at $\approx1280~\muHz$, which were then distributed to several modeling teams. Using different approaches and input physics each team returned estimates of the physical properties of the star that were consistent to within $1-2\sigma$, suggesting that the application of asteroseismic constraints produces more robust estimates of the stellar properties. For the mass and radius, which are typically well constrained by asteroseismology \citep{Lebreton2014, Stokholm2019}, we adopted the overall values $\Mstar=1.16\pm 0.03~\Msun$ and $\Rstar=1.63\pm 0.04~\Rsun$. Together with a surface temperature of $\Teff = 5829\pm80$~K. Previous literature estimates suggest \lamfor could be anything from an early G dwarf to a late G-type subgiant; however, multiple modeling teams using asteroseismic constraints all place the star firmly at the start of the subgiant phase of its evolution.

The age of the system was less well constrained, despite the seismic constraint, with an estimate of $6.3\pm0.9$~Gyr, with most literature values falling within this range. This is likely due to the correlation with the other model parameters, where particularly the mass and metallicity are important for estimating the age. In our case the uncertainty on the mass estimate is caused in part by the uncertainty on the mode frequencies due to the relativity short \tess time series, while the metallicity is taken from spectroscopic values in the literature. 

Following this we revisited the analysis of \lamforb originally performed by \citet{Otoole2009}, who discovered the Neptune-like planet. We combined the radial velocity measurements from the original publication with the now public HARPS measurements, yielding an RV time series spanning approximately 20 years. Combining this with the asteroseismic mass estimate, reduces the lower mass limit of \lamforb from $m_p\sin{i}=22.1\pm2.0~\Mearth$ to $m_p\sin{i}=16.8^{+1.2}_{-1.3}~\Mearth$. The majority of this reduction is due to the long RV time series, which alone sets a lower limit of $m_p \sin{i}=17.2\pm1.3~\Mearth$. The stellar mass found here using asteroseismology is $\sim3\%$ lower than that used by \citet{Otoole2009}, and as such reduces the lower limit on the planet by a similar amount.

The orbital eccentricity was also found to be significantly higher at $e=0.35\pm0.05$, as opposed to the previous estimate of $e=0.20\pm0.09$. We estimate the circularization timescale due to tidal interaction following an expression derived in \citet{Barker2009} and find it to be $\approx1120$~Gyr, which is much longer than the age of the system. The slightly higher eccentricity found here is then perhaps more consistent with this long circularization timescale, compared to the original estimate, which at a $2\sigma$ level encompasses almost circular orbits.

In  addition to the revised parameters for \lamforb, the larger set of RV measurements also revealed a periodicity at $33$ days. Despite the relatively low activity level of the star, this periodicity is more likely due to stellar rotation when compared to the case of an additional unknown planet being present in the system. Although difficult to confirm, the former scenario is consistent with the expected rotation rate of an old inactive star like \lamfor, and we showed that the latter scenario is not possible as such an orbit would be unstable after $\sim10^3$ years. Assuming then that the $33$-day RV signal is indeed due to rotation, the relatively short $17.25$ day orbit of \lamforb means that tidal interaction with the host star will cause the planet to gradually spiral inward into the star. Using the relation by \citet{Barker2009} we can estimate the current infall timescale to be on the order of $10^3-10^4$~Gyr. This is obviously much longer than the evolutionary timescale of the host star, and so the time when the star expands to the current periastron of \lamforb (0.1~AU, $21.5~\Rsun$) sets an upper limit for when the planet will be engulfed. The models presented in Section \ref{sec:stellarmodeling} suggest that this will happen in approximately $1.5$~Gyr. However, the tidal interaction, and thus the in-fall timescale, is a strong function of the stellar radius and the orbital period \citep{Barker2009}. This will cause the in-fall to accelerate considerably over time and the planet will likely be engulfed well before \lamfor expands to the current orbit.

Despite the rather modest time series that was obtained from \tess for \lamfor, we have shown that it is still possible to measure the individual oscillation frequencies of the star. Moreover, using these frequencies, multiple modeling teams find consistent results on the percent level for the mass and radius of \lamfor. This shows that asteroseismology is a useful tool for obtaining robust constraints on these stellar parameters, for the enormous selection of stars being observed by \tess, which previously has only been possible for select areas of the sky like those observed by \corot and \kepler. This new and much wider range of stars that \tess is observing prompts the reexamination of the wealth of archival radial velocity data that has been accumulated in the last few decades for planet hosting systems, in order to better characterize these systems.

\begin{acknowledgements}
The authors would like to thank J. P. Faria and H. Rein for useful discussions. This paper includes data collected by the TESS mission. MBN, WHB, MRS, AHMJT, and WJC acknowledge support from the UK Space Agency. AHMJT and MRS have benefited from funding from the European Research Council (ERC) under the European Union's Horizon 2020 research and innovation programme (grant agreement n${^\circ}$ 803193/BEBOP). Funding for the Stellar Astrophysics Centre is funded by the Danish National Research Foundation (Grant agreement no.: DNRF106). ZÇO, MY, and SÖ acknowledge the Scientific and Technological Research Council of Turkey (T\"UB\.ITAK:118F352). AS acknowledges support from grants ESP2017-82674-R (MICINN) and 2017-SGR-1131 (Generalitat Catalunya). TLC acknowledges support from the European Union's Horizon 2020 research and innovation programme under the Marie Sk\l{}odowska-Curie grant agreement No.~792848 (PULSATION). This work was supported by FCT/MCTES through national funds (UID/FIS/04434/2019). MD is supported by FCT/MCTES through national funds (PIDDAC) by this grant UID/FIS/04434/2019. MD and MV are supported by FEDER - Fundo Europeu de Desenvolvimento Regional through COMPETE2020 - Programa Operacional Competitividade e Internacionalizaç\~ao by these grants: UID/FIS/04434/2019; PTDC/FIS-AST/30389/2017 \& POCI-01-0145-FEDER-030389 \& POCI-01-0145-FEDER03038. MD is supported in the form of a work contract funded by national funds through Fundação para a Ciência e Tecnologia (FCT). SM acknowledges support by the Spanish Ministry with the Ramon y Cajal fellowship number RYC-2015-17697. BM and RAG acknowledge the support of the CNES/PLATO grant. DLB and LC acknowledge support from the TESS GI Program under awards 80NSSC18K1585 and 80NSSC19K0385. LGC thanks the support from grant FPI-SO from the Spanish Ministry of Economy and Competitiveness (MINECO) (research project SEV-2015-0548-17-2 and predoctoral contract BES-2017-082610). Funding for the TESS mission is provided by the NASA Explorer Program. Based in part on data acquired at the Anglo-Australian Telescope. We acknowledge the traditional owners of the land on which the AAT stands, the Gamilaraay people, and pay our respects to elders past and present. The data presented herein were in part obtained at the W. M. Keck Observatory, which is operated as a scientific partnership among the California Institute of Technology, the University of California and the National Aeronautics and Space Administration. The Observatory was made possible by the generous financial support of the W. M. Keck Foundation. The authors wish to recognize and acknowledge the very significant cultural role and reverence that the summit of Maunakea has always had within the indigenous Hawaiian community. 
\end{acknowledgements}

\bibliographystyle{aa}
\bibliography{main}

\begin{appendix}
\section{Manual time series reduction}
\label{app:manual}
For each \tess orbit we extracted a time series for each pixel and took the brightest pixel as our initial time series. The pixel time series quality figure of merit was parameterized by ${q = \sum_{i=1}^{N-1}\mid f_{i+1}-f_{i}\mid}$, where $f$ is the flux at cadence $i$, and $N$ is the length of the time series. Using the first differences of the light curve acts to whiten the time series, and thus correct for its non-stationary nature \citep{Nason2006}; similar approaches have been used in astronomical time series analysis by \citet{Buzasi2015} and \citet{Prsa2019}, among others. 

We then iteratively added the flux of the pixels surrounding the brightest pixel. The process continued until the light curve quality stopped improving, and the resulting pixel collection was adopted as our aperture mask. The light curve produced by our aperture mask was then detrended against the centroid pixel coordinates by fitting a second-order polynomial with cross terms. Similar approaches have been used for K2 data reduction \citep[see, e.g.,][]{Vanderburg2014}.

\begin{figure}
    \centering
    \includegraphics[width = \columnwidth]{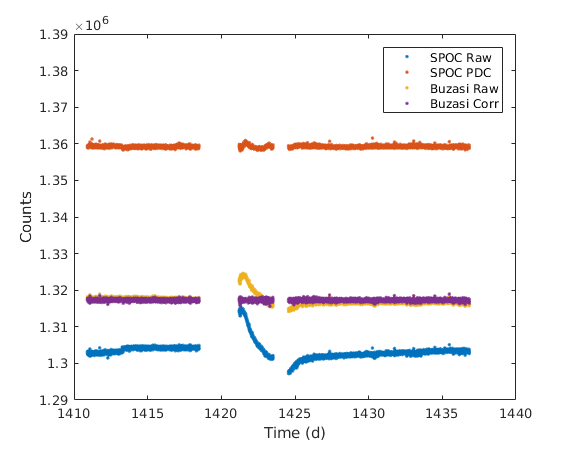}
    \caption{Resulting counts over time for different \tess data reduction pipelines. The Buzasi Corr time series very effectively removes most of the long-period variability; however, the SPOC time series still shows the lowest variance in the frequency range around the \pmode envelope. 
    \label{fig:LC_comparison}}
\end{figure}

Figure \ref{fig:LC_comparison} shows the resulting time series, compared to that derived by the SPOC. In this case, low-frequency noise was somewhat improved over the SPOC light curve product, but noise levels at the frequencies near the stellar oscillation spectrum were not. We accordingly used the SPOC light curve for the analysis outlined above.

\section{Peakbagging frequencies}
\label{app:pb}
\begin{table}[]
    \centering
    \caption{Oscillation frequencies $\nu$ with angular degree $l$ of \lamfor}
    \onehalfspacing
    \begin{tabular}{cc}
        \hline
        $l$ & $\nu$ [$\muHz$] \\
        \hline
        $0$ & $1142.23^{+2.08}_{-2.00}$\\
        $1$ & $1171.79^{+1.55}_{-1.41}$\\
        $2$ & $1206.95^{+1.96}_{-2.13}$\\
        $0$ & $1211.30^{+0.44}_{-0.69}$\\
        $1$ & $1243.36^{+0.80}_{-0.92}$\\
        $0$ & $1282.05^{+1.83}_{-1.98}$\\
        $1$ & $1312.27^{+1.29}_{-0.97}$\\
        $2$ & $1348.67^{+1.81}_{-2.70}$\\
        $0$ & $1351.17^{+0.43}_{-0.44}$\\
        $1$ & $1380.81^{+0.45}_{-0.58}$\\[1ex]
        \hline
    \end{tabular}
   \tablefoot{The frequency resolution of the data set is $\Delta T^{-1}=0.2067\muHz$}

    \label{tab:freqs}
\end{table}

\section{Posterior distributions}
\begin{figure}
    \centering
    \includegraphics[width = 1\columnwidth]{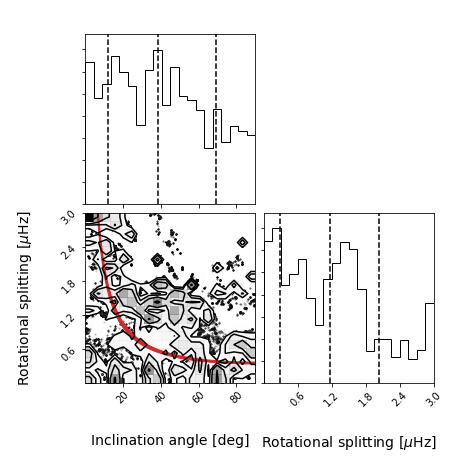}
    \caption{Corner plot of the rotational splitting and inclination angle posterior distributions from the seismic fit, consisting of $10^5$ samples. The marginalized posterior distributions are shown in the diagonal frames. The dashed vertical lines show the $16$th, $50$th, and $84$th percentiles of the distributions at $i=38^{+31}_{-26}$ degrees and $\nu_{rot}=1.2\pm0.9\muHz$ for the inclination and rotational splitting respectively. The lower left frame shows a 2D histogram of the distributions (black). The shaded red curve shows the rotational splittings corresponding to $v\sin{i} = 2.5\pm0.1$km/s from \citet{AmmlerVonEiff2012} and a stellar radius of $\Rstar=1.63\pm0.04\Rsun$.}
    \label{fig:banana}
\end{figure}

\begin{figure}
    \centering
    \includegraphics[width = 1\columnwidth]{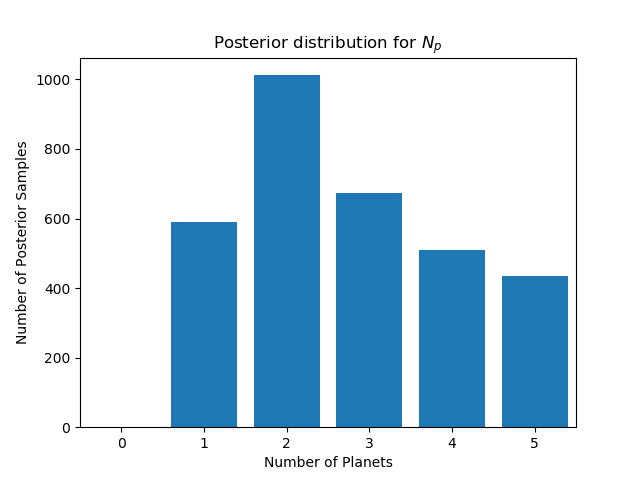}
    \caption{Marginalized posterior of the $N_p$ parameter in the \kima fit to the RV measurements of \lamfor. The number of samples in each bin corresponds to the likelihood, while the ratio of the height of each bin indicates the Bayes factor of one configuration over another. Comparing the cases of $N_p=0$ and $N_p=1$ the Bayes factor is effectively infinite, indicating that there is  at least one planet in the system. In contrast, the ratio between $N_p=2$ and $N_p=1$ is low, with a Bayes factor of $1.67$, suggesting that there is little evidence to support a two-planet configuration.}
    \label{fig:np}
\end{figure}

\begin{figure*}
    \centering
    \includegraphics[width = \columnwidth]{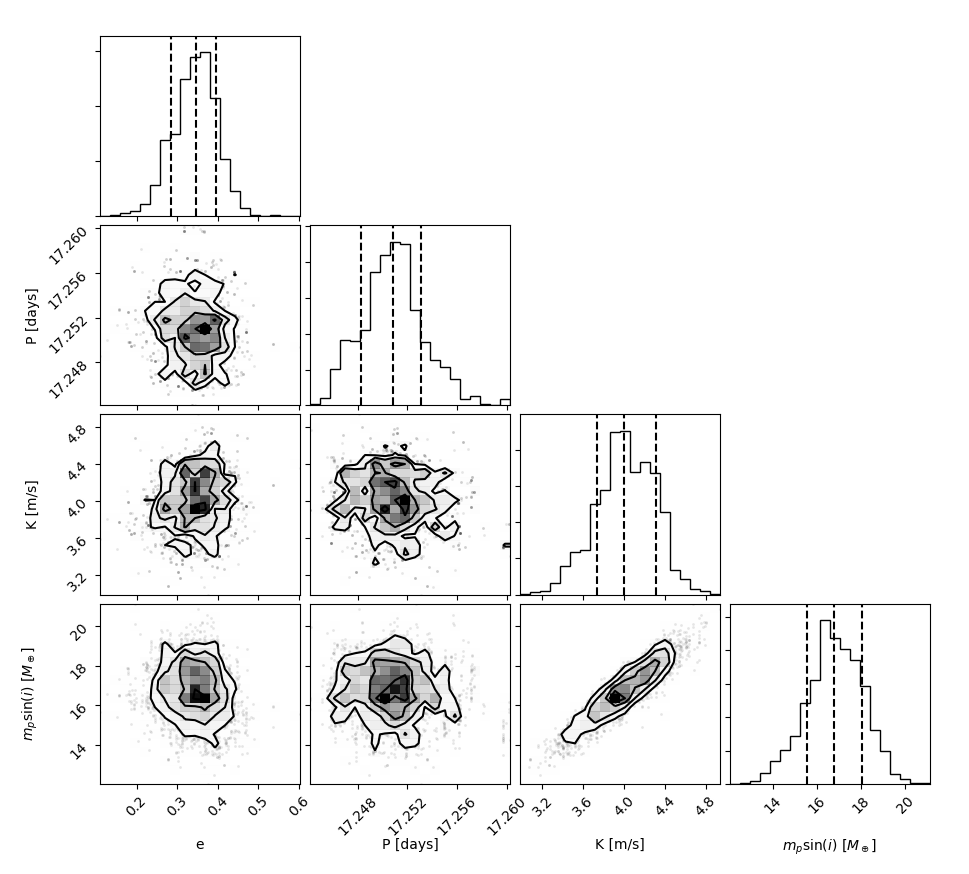}
    \caption{Corner plot of the orbital parameters from the \kima fit: the eccentricity $e$, the orbital period $P$, velocity semi-amplitude $K$, and the projected planet mass $M$.}
    \label{fig:planetcorner}
\end{figure*}

\section{Stellar properties literature values}

\begin{table}[]
    \centering
    \caption{Literature sources for $\Teff$ and $\logg$ values shown in Fig. \ref{fig:comparison}.}
    \onehalfspacing
    \begin{tabular}{lll}
        \hline
        Source & $\Teff$ [K] & $\logg$ [cm/s$^2$]\\
        \hline
        \citet{Hearnshaw1972}  & $5793$ & $4.09$ \\
        \citet{Gehren1981}  & $5860$ & $4.35$ \\
        \citet{Bensby2003}  & $5800\pm70$ & $4.04\pm0.1$ \\
        \citet{Valenti2005} & $5817\pm44$ & $4.17\pm0.06$ \\
        \citet{daSilva2006}  & $5936\pm70$ & $4.12$ \\
        \citet{Gray2006} & $5745$ & $4.11$ \\
        \citet{Bond2006}  & $5374\pm57$ & $4.05\pm0.19$ \\
        \citet{Sousa2006}  & $5876\pm22$ & $4.22\pm0.01$ \\
        \citet{Sousa2008}  & $5841\pm17$ & $4.16\pm0.02$ \\
        \citet{Tsantaki2013}  & $5843\pm12$ & $4.16\pm0.03$ \\
        \citet{Carretta2013}  & $5821$ & $4.1$ \\
        \citet{Ramirez2014}  & $5817\pm15$ & $4.146\pm0.024$ \\
        \citet{Bensby2014}  & $5885\pm72$ & $4.23\pm0.1$ \\
        \citet{Datson2015}  & $5766\pm40$ & $4.16\pm0.07$ \\
        \citet{BertrandeLis2015}  & $5841$ & $4.16$ \\
        \citet{Battistini2015}  & $5885$ & $4.2$ \\
        \citet{Bonfanti2016}  & $5818$ & $4.06\pm0.02$ \\
        \citet{DelgadoMena2017} & $5841\pm17$ & $4.16\pm0.02$ \\
        \hline

    \end{tabular}
    \label{tab:lit}
\end{table}

\section{HARPS observing programs}
\label{app:progID}
\begin{table}[]
    \centering
    \caption{HARPS observing program PIs and IDs for data used in this work.}
    \onehalfspacing
    \begin{tabular}{ll}
        \hline
        PI & Program ID \\
        \hline
        Diaz & 198.C-0836 \\
        Doellinger & 078.C-0751 \\
        Doellinger & 079.C-0657 \\
        Doellinger & 081.C-0802 \\
        Doellinger & 082.C-0427 \\
        Hatzes & 074.C-0102 \\
        Mayor & 072.C-0488 \\
        Udry & 091.C-0936 \\
        Udry & 183.C-0972 \\
        Udry & 192.C-0852\\
        \hline
    \end{tabular}
    \label{tab:progID}
\end{table}

\end{appendix}
\end{document}